\documentclass[twocolumn]{aastex631}
\usepackage{amsmath}
\usepackage{wrapfig}

\begin{document}

\title{Starspots and Flares are Generally Not Correlated}

\author[0009-0005-6169-6413]{Andy B. Zhang}
\correspondingauthor{Andy.Zhang649964@tufts.edu}
\affiliation{Department of Physics \& Astronomy, Tufts Astronomy, 574 Boston Avenue, Medford, MA, USA}

\author[:0009-0002-9757-0351]{Jason R. Reeves}
\affiliation{Department of Physics \& Astronomy, Tufts Astronomy, 574 Boston Avenue, Medford, MA, USA}

\author[0000-0002-7595-6360]{David V. Martin}
\affiliation{Department of Physics \& Astronomy, Tufts Astronomy, 574 Boston Avenue, Medford, MA, USA}

\author[0009-0001-8728-6894]{Veronica Pratt}
\affiliation{Department of Physics \& Astronomy, Tufts Astronomy, 574 Boston Avenue, Medford, MA, USA}

\author[0009-0001-8728-6894]{Wata Tubthong}
\affiliation{Department of Physics \& Astronomy, Tufts Astronomy, 574 Boston Avenue, Medford, MA, USA}

\author[0000-0000-0000-0000]{Arielle Weinstein}
\affiliation{Department of Physics \& Astronomy, Tufts Astronomy, 574 Boston Avenue, Medford, MA, USA}

\author[0000-0000-0000-0000]{Isabella E. Ward}
\affiliation{Department of Physics \& Astronomy, Tufts Astronomy, 574 Boston Avenue, Medford, MA, USA}



\begin{abstract}

Sunspots and solar flares are two different manifestations of magnetic activity on the surface of the Sun. On the Sun, flares typically occur close to spots. In this paper we test this the connection between spots and flares on other stars. We detect 218,386 stellar flares on 14,163 spotted stars using a new algorithm called \textsc{toffee}. Inhomogeneous spot distributions mean that as stars rotate they become brighter when less spots are facing the observer, and dimmer when more spots are facing the observer. We determine that flares occur when the star is brighter $49.97\pm 0.21\%$ of the time, i.e. there is an equal preference for the flares to occur when the star is relatively bright or dim. We therefore find no evidence for a correlation between flare rate and spot occurrence, contrary to what is seen on the Sun.


\end{abstract}



\section{INTRODUCTION}\label{sec:introduction}

Stellar flares and starspots are both manifestations of magnetic activity generated by the stellar dynamo. \citep{2020charbonneau}. Starspots are relatively cold regions on the stellar surface which form where strong magnetic flux concentrations inhibit convective heat transport in the photosphere. Flares arise from the rapid release of magnetic energy through reconnection in the overlying corona. These phenomena are among the most visible signatures of stellar activity and have important implications for our understanding of stellar and solar astrophysics. There are also implications for the  environment and habitability of orbiting planets.

The majority of solar flares originate within or immediately adjacent to magnetically active regions that host sunspots \citep{Fletcher2011}. This trend is particularly strong for moderate-energy C-class and M-class flares, as well as for the most energetic X-class events \citep{2014Guo,Li2023,2023Oloketuyi}. In contrast, weak B-class flares occur more frequently in regions with little or no visible spot coverage. Nevertheless, both the total number of flares and, especially, the total flare energy budget are dominated by events arising from sunspot-bearing active regions.

Conducting a comparable study on other stars is impeded by limited spatial information about the spots and flares. Techniques to map spot locations do exist, including lightcurve inversion \citep{Walkowicz2013}, Doppler imaging \citep{Vogt1983,Strassmeier1998}, interferometry \citep{Roettenbacher2017} and spot crossings from transiting planets \citep{Nutzman2011,Silvavalio2011}. All of these methods have challenges that inhibit them from being applied to a large sample of stars. It is even more difficult to spatially locate stellar flares, with only sporadic attempts at this made in the literature, including lightcurve inversion \citet{Ilin2021} and flare occultations from transiting bodies \citep{2024Martin,Armitage2025}.

To date, studies of the correlation between star spots and flares have focused on the distribution of flares as a function of rotation phase. This has been made possible by long baseline, high precision photometric campaigns like TESS and Kepler. \citet{HuntWalker2012,2014Hawley,2015Lurie,2018Doyle,2019Doyle,2020Doyle} all searched for flares on small samples of spotted stars and discovered that flares were uniformly distributed in rotation phase. \cite{2018Roettenbacher} used a larger sample of  2,447 flares on 111 stars and found a slightly non-uniform distribution. They found a small correlation between low-amplitude flares ($1-5\%$ flux increase) and spot minimum flux, but no such correlation for higher amplitude flares.

Most past studies have considered flares as a function of the rotational phase of the star, as seen in the lightcurve. This method is challenging because the evolution of spots and the differential rotation of stars mean that rotation phase is difficult to define. This also makes it challenging to combine results across multiple stars; most past studies have focused on individual stars. Furthermore, to study a correlation between flare occurrence and spot phase, one needs a lot of flares across all phases to make a statistically strong statement. 

Rather than study flares as a function of rotational phase, one can study flares as a function of spot modulation flux. Simply put, the problem is broken down into a binary question: do more flares occur when a star is brighter (less spots rotated into view) or dimmer (more spots rotated into view) than the star's median flux? \cite{2024Martin} applied this to the double M-dwarf eclipsing binary CM Draconis. They found a slight increase in flare rate with positive spot flux, i.e. flares were more likely to occur when the star was seemingly less spotty. However, this was only based on 163 flares. The most recent TESS data has contributed over 100 more flares and this correlation has seemingly disappeared (priv. comm.). Furthermore, since CM Dra contains nearly identical stars, it is essentially impossible to assign spots and flares to a given star.

In this paper we study flare occurrence as a function of spot modulation amplitude. Our methodology builds upon that used in \citet{2024Martin}, but we improve it in several ways: over 14,000 stars instead of 1, an avoidance of eclipsing binaries, and over 200,000 flares instead of 163. Our flare detection is improved through the implementation of our new algorithm \textsc{toffee}, which was first applied in Pratt et al. (under rev.) in the study of sympathetic flares, and is publicly available on Github\footnote{\url{https://github.com/JasonReeves702/TOFFEE}}. Our spot modeling is also significantly improved relative to \citet{2024Martin}, allowing us to handle more complex and variable spot morphologies.

Our paper is organized as follows.  Sections~\ref{sec:flare_detection_and_modeling} and ~\ref{sec:spot_modeling} show how we find flares and spots, respectively. In Sect.~\ref{sec:bias} we demonstrate our process for filtering the sample of stars and flares to make our results as unbiased as possible. In Sect.~\ref{sec:results} we present our results and discuss a few possible caveats and complications, before concluding in Sect.~\ref{sec:conclusion}.


\section{Flare Detection and Modeling} \label{sec:flare_detection_and_modeling}

\subsection{Stellar Sample} \label{sec:stellar_sample}

This paper uses stellar samples collected from \cite{2024Feinstein}, \cite{2025Yudovich}, and \cite{2025seli}. Using samples previously considered for past stellar flaring papers provides a frame of reference on the performance of our flare detection algorithm in picking up temporally close and distant flaring events. For all TESS data, we use 120 second cadence data from Pre-search Data Conditioning Simple Aperture Photometry (PDCSAP) flux lightcurves after performing a quality cut to only consider points with a quality flag of zero. 



For the sample adopted from \cite{2024Feinstein} we use the 3,160 star subsample they found to have flaring events between TESS sectors 1 to 67 for an equal comparison for our flare detection algorithms in the same 9,274 sector lightcurves. For the sample from \cite{2025Yudovich} we adopt their sample of 234 M-class dwarfs originally found as a subset as the stellar sample from \cite{2022Crowley}. We use the 1,902 lightcurves for the sample of stars from TESS cycles 1-4 (sectors 1-55). For the sample adopted from \cite{2025seli}, we adopt their sample of 14,408 stars from across the main sequence. Limiting to lightcurves from TESS sectors 1-69, we use a total of 58,406 lightcurves from the \cite{2025seli} sample. When accounting for the overlap between the samples, this study considers a total of 16,305 unique stellar targets and 62,450 unique lightcurves. Note that our final stellar sample will be smaller than this, since we require not just detectable flares but visible star spots. We also apply a procedure of cuts to the star and flare samples to minimize bias in our final results.

\subsection{Lightcurve Detrending} \label{sec:detrending}

To discover the flares, we need to remove any non-flare instrumental and astrophysical trends, particularly the effects of spot modulation and stellar pulsation. We segmented each light curve by TESS orbit, fitting and dividing by a quadratic trend using \textsc{Numpy}'s polyfit function to remove any instrumental trends on the timescale of a TESS orbit (13.7 days). We then utilized Tukey's biweight filter \citep{1977Tukey} provided by the \textsc{Wotan} package \citep{2019Hippke} applied to our light curves with a standard $0.25$ day window length, a threshold below which we observed noticeable attenuation of flares by the flattening algorithm, harming potential detection.

\begin{figure}
    \centering
    \includegraphics[width=0.75\linewidth]{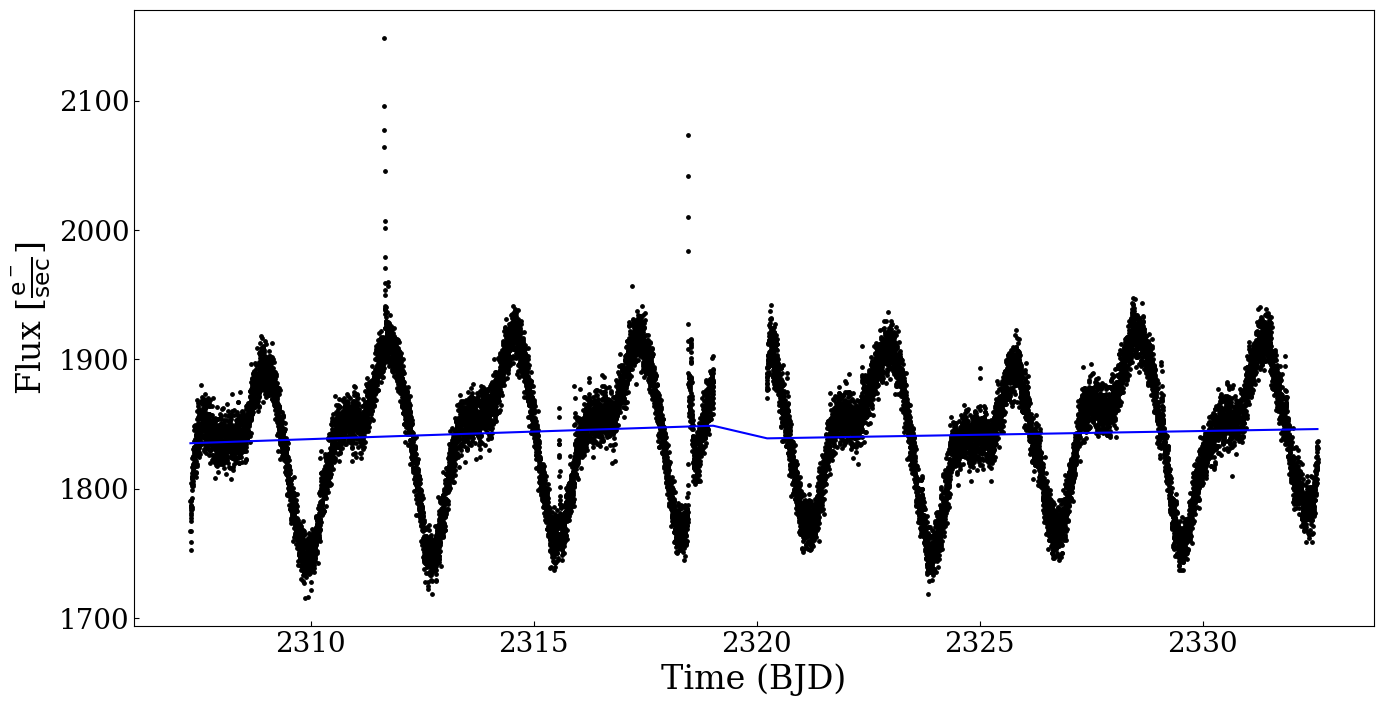}
    \includegraphics[width=0.75\linewidth]{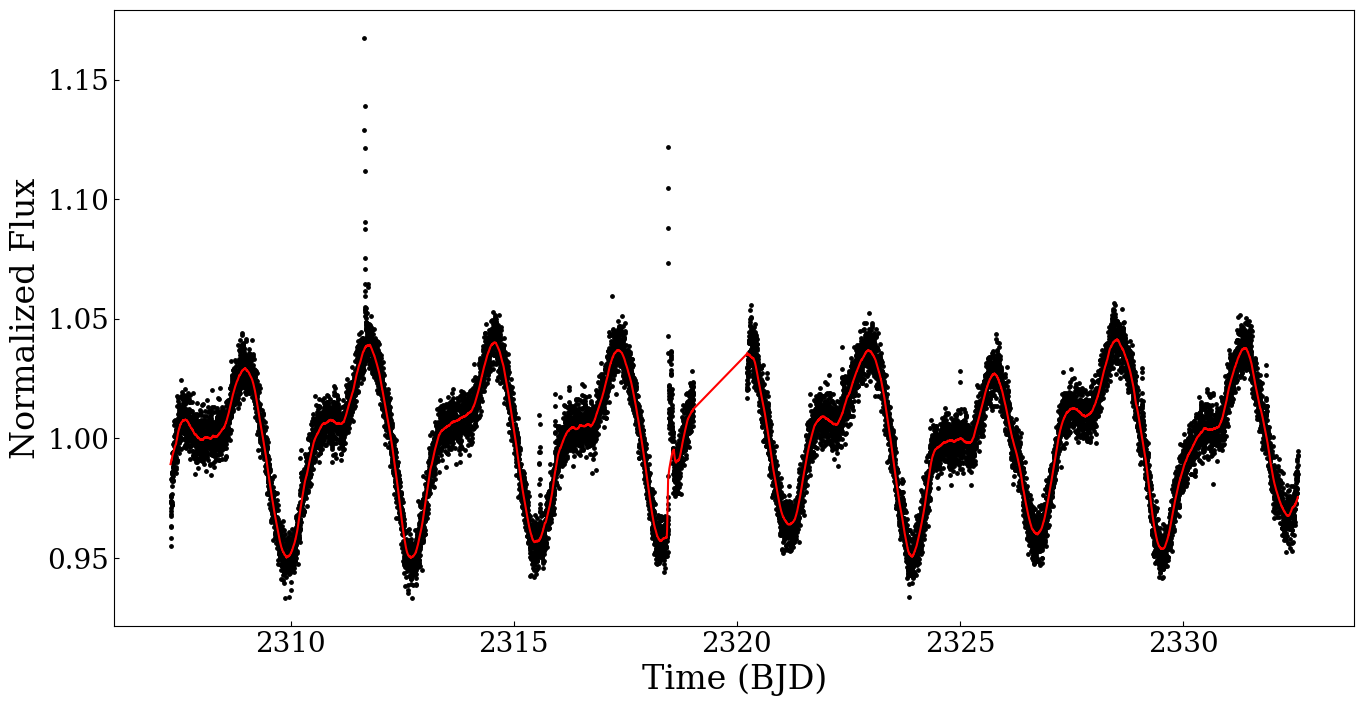}
    \includegraphics[width=0.75\linewidth]{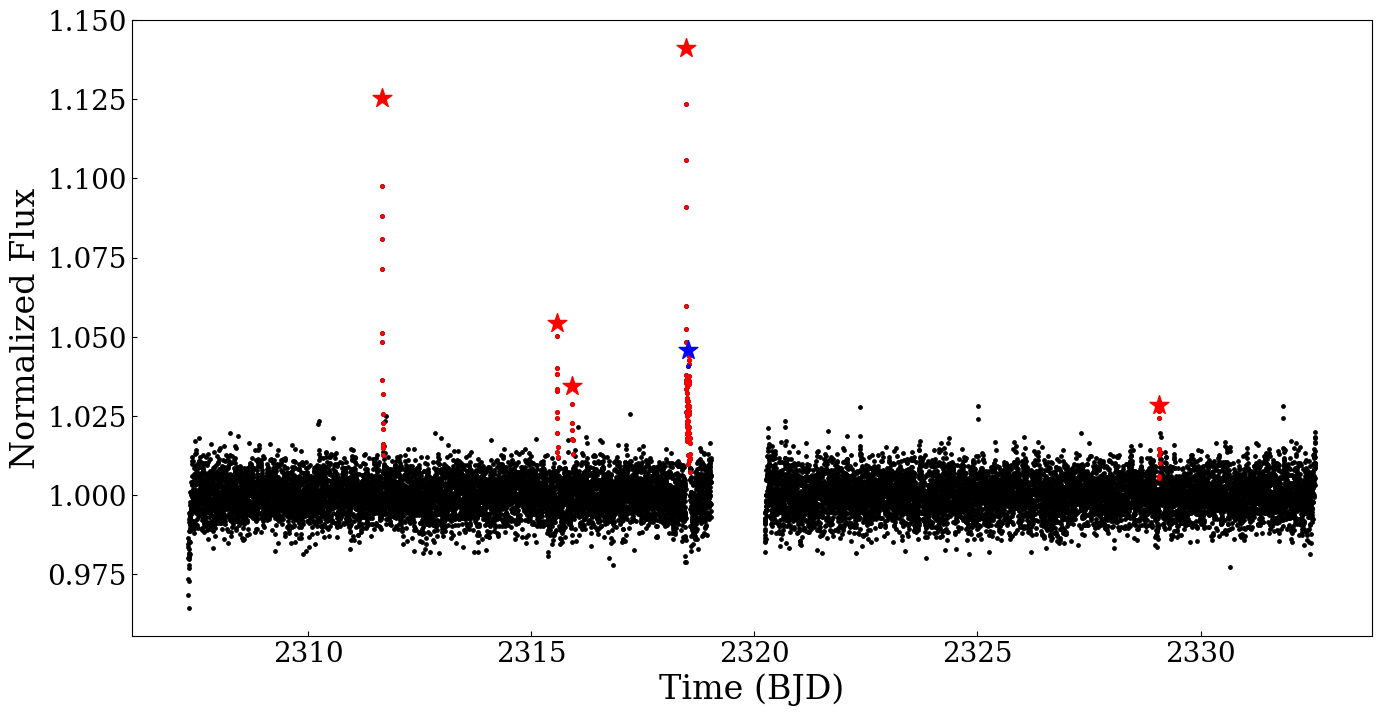}
    \caption{Detrending and flare search process of Sector 37 of BY Draconis variable star TIC 378126824 from raw lightcurve (\textit{top}) to final flattened curve (\textit{bottom}). \textit{Top} shows the raw time resolved brightness of the star in black points with the quadratic trend coming from the orbit of the TESS telescope in blue. \textit{Middle} shows the quadratic subtracted lightcurve overlaid with the trend found by \textsc{Wotan} in red. The bottom curve is the final flattened lightcurve in black and the detected flares. Primary flares are colored in red with the peaks labeled as red stars and the secondary is labeled in blue with a blue star representing its peak.}
    \label{fig:detrending}
\end{figure}

Next, a Lomb-Scargle periodogram was created per TESS orbit of data. If the false alarm probability (FAP) of the identified peak periodicity relative to Gaussian noise was above $20\%$, we deemed there to be no remaining periodicity, and the light curve to be sufficiently flattened. Otherwise, we conducted further flattening with \textsc{Wotan}'s `median' filter on a window length equal to half the periodogram's peak signal period. This procedure is then repeated iteratively on the new light curve until the FAP has reached above $20\%$, adjusting window size to the residual periodicity of each light curve through each step. Lastly, we applied a mask to each lightcurve after detrending to cut off 100 points on either side of each break. Fig. \ref{fig:detrending} demonstrates the process from taking a raw lightcurve, through the bi-weight flattening, before labeling the detected flares using \textsc{toffee}.

We note that due to this iterative flattening at window lengths below $0.25$ days, the amplitudes of flares are attenuated and are not representative of their true normalized fluxes, particularly for higher amplitude, long duration flares. Nonetheless, this process preserves the amplitude hierarchy between any two given flares and has minimal impact on the amplitude – and thus detectability – of low amplitude flares. 

\subsection{Flare Detection} \label{sec:flare_detection}

We utilized the threshold-based flare detection algorithm \textsc{toffee} as outlined in Pratt et al. (under rev.) and Reeves et al. (under rev.). This method takes detrended light-curves and detects epochs of increased emission as flaring events. This method is similar to other threshold-based algorithms outlined in \cite{2021Ilin, 2014Davenport, 2019Yang, 2022Howard} but features additional functionality to find secondary flares and limit false-positive detection from spot modulation.

Like other threshold-based methods, \textsc{toffee} finds bright flux points above a given flux threshold compared to the global spread across the lightcurve. However, unlike other algorithms, this code does not strictly scan for a certain number of points consecutively above a given sigma threshold to label as a flare. Instead, \textsc{toffee} works by finding all the points above the flux threshold then works down from the brightest points to the least bright above the threshold, modeling around each point determining to which specific flaring events the bright points potentially belong to. Because the standard deviation can be affected by outliers such as numerous bright flares, we instead use percentile rank to better represent the spread of the flux due to Poisson noise and residual spot modulation. The median of the post-flattened lightcurve is subtracted from the flux to center around zero, then the 84-th percentile rank value was taken to represent one standard deviation above the typical quiescent flux, which is then multiplied by three to calculate the $3\sigma$ threshold for detection. We used the spread of the flattened lightcurves rather than the photometric error of each flux point to account for possible residual spot modulation, inflating the errors to require the fluxes to be slightly larger in order to be counted as a flare.

In order to better capture low amplitude flares, we relaxed the requirement used in \cite{2021Ilin} and \cite{2015Chang} where bright points need to be consecutive. Instead, we required that the possible flaring event have three in four flux measurements above the $3\sigma$ limit. The concern of allowing few cadences to determine flare detection is possible contamination from cosmic rays, however ultimately requiring three points above the flux threshold makes a robust case for any given signal being a flare. Such a requirement also is seen to be effective in eliminating false-positive detrending artifacts resulting from spot modulation where epochs of the lightcurve rise and fluctuate around the $3\sigma$ threshold but fail to have 75\% of such points above the flux threshold. To counter the potential side-effect where a true flare occurs on the rise of residual red noise such that the epoch of the lightcurve surrounding the flare fluctuates around the threshold, we added an additional check that if there are three consecutive points above $4.5\sigma$, the epoch is still considered a valid flare detection. Before moving to the next potential flaring event, 

\textsc{toffee} models the flare with a \citet{2014Davenport} double-exponential model. This model is subtracted and then \textsc{toffee} is run again on the residuals to search for additional impulses within the original flare to find secondary flares, i.e. overlapping flares (see Pratt et al., under rev. for additional detail). Here, \textsc{toffee} is capable of finding secondary flare events with amplitudes as low as $2\sigma$. An epoch that passes this check is thus declared a flare and added to the sample. All other bright points that were previously identified in the first step but determined to be a part of this flare are thrown out. \textsc{toffee} records the flare start and end times, flare peak time, amplitude, equivalent duration, and a series of flags on how the flare was detected.


\section{Spot Modeling} \label{sec:spot_modeling}

\subsection{Spot Model} \label{sec:double_D_sine}
To ensure a sufficiently accurate modeling of the local light curve variations, we choose model the flux $f$ as a function of time $t$ with a double sinusoidal function, where for each sine  function the amplitude can vary quadratically in time:
\begin{multline}
f= (At^2+Bt+C)* \sin(Dt+E) \\
+ (Ft^2+Gt+H)* \sin(It+J) + K \ .          
\end{multline}

The reason for two sine curves is that many light curves, particularly M-dwarfs, show a double-humped feature, as is common in so-called BY Draconis variables. AU mic is a particularly strong example of this double humped nature \citep{Gilbert2022}.Double-humped spot modulation is also common in tight eclipsing stellar binaries \citep{Sethi2024}, although we avoid eclipsing binaries in this paper by cutting all flares with poor model fits, which include those due to transits/eclipses (see Sect. \ref{sec:poor_spot_mod}). The quadratic amplitude variation accounts for both spot evolution and differential rotation. 

Overall, this model provides the flexibility needed to consistently trace out sinusoidal modulations to the light curve of varying complexity, without being susceptible to overfitting and oversensitivity to localized variance in the lightcurves.

\subsection{Fit Criteria} \label{sec:fit_criteria}
To properly fit for the spot modulation, we first mask out all data during the flares discovered by \textsc{toffee}. To account for incomplete capture of the flare rise and decay by \textsc{toffee}, we also additionally mask any data within 60 minutes or two flare durations of the flare's end as labeled by \textsc{toffee}, whichever is longer. Similarly, data within half the flare duration of the flare's start is also masked. The length of these added data cuts was chosen such that the fitted spot amplitudes of a set of injected flares on synthetic light curves demonstrated no bias to either positive or negative spot modulation, as described in sect. \ref{sec:synth_flares}.

\begin{figure}
    \centering
    \includegraphics[width=1\linewidth]{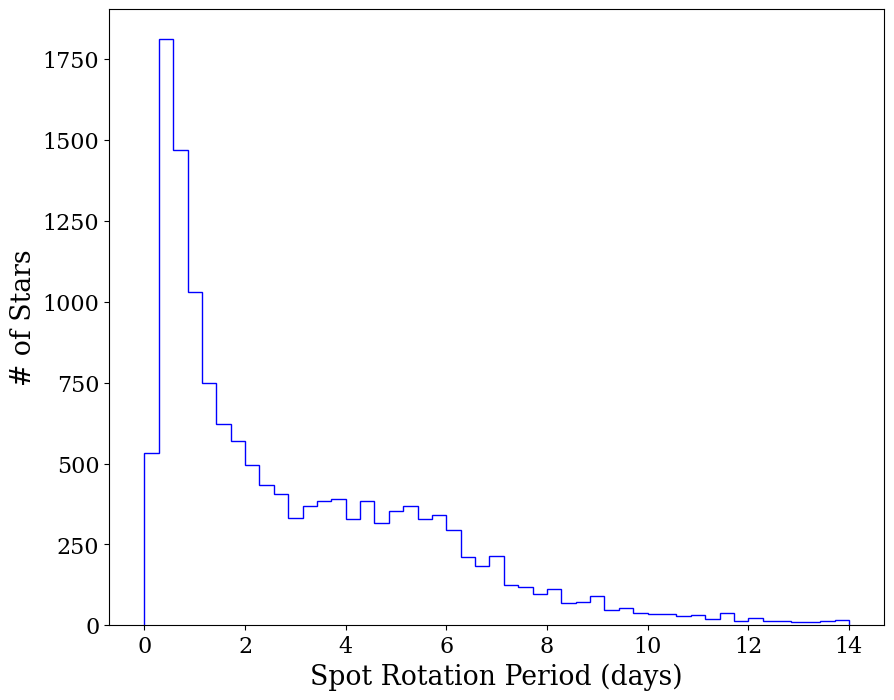}
    \caption{Rotation periods of 14,163 stars with a Lomb-Scargle periodogram false alarm rate less than 5\%.}
    \label{fig:rot_periods}
\end{figure}

We then create Lomb-Scargle periodograms for each light curve, similar to the flattening procedure for flare detection in \ref{sec:detrending}, here over all TESS sectors and after the per-sector quadratic trends have been removed, but prior to any flattening with \textsc{Wotan} has been conducted to identify the general periodicity in full. Of the 16,410 \textsc{toffee} processed stars across the three samples, we find 14,163 containing sinusoidal signals with less than a 5\% false alarm rate, shown in Fig. \ref{fig:rot_periods}. The median rotation period is 2.16 days. 

The peak signal period of the global periodogram is then used to define a four-period wide window for each flare, centered around the flare's peak amplitude time. If there are any gaps in this data at least $25\%$ of the global periodogram period wide, the window is further shortened to avoid them. A local Lomb-Scargle Periodogram is then run on the remaining data within the window to determine the local periodicity. If the False Alarm Probability (FAP) is lower than $5\%$, the algorithm moves on to spot modulation fitting for the flare.

\subsection{Fit Procedure} \label{sec:fit_procedure}
The data within the window is now fitted to the above sinusoidal model with \textsc{Scipy}'s curve fit function, with two different sets of initial parameters. Both sets have constant amplitudes taken from the window data. They differ only by the frequency coefficient of the secondary sinusoid, wherein one is double the determined primary frequency and the other is half. The better of the two fits is then chosen using the reduced chi-squared statistic, and the spot modulation flux at flare peak time is interpolated from the model. The amplitude of the spot when a given flare occurs is then calculated from the equation below:

$$
\text{Spot Amplitude}=2*\frac{f(t_{peak})-\min{f(t)}}{\max{f(t)} - \min{f(t)}} - 1
$$

where $f(t)$ is the fitted spot modulation flux for times $t$ in the local window, and $t_{peak}$ is the \textsc{toffee}-identified peak flare time. The resulting value will be between $-1 ~\text{and} ~1$, with the parity corresponding to whether the fit is positive or negative.

\begin{figure}
    \centering
    \includegraphics[width=1\linewidth]{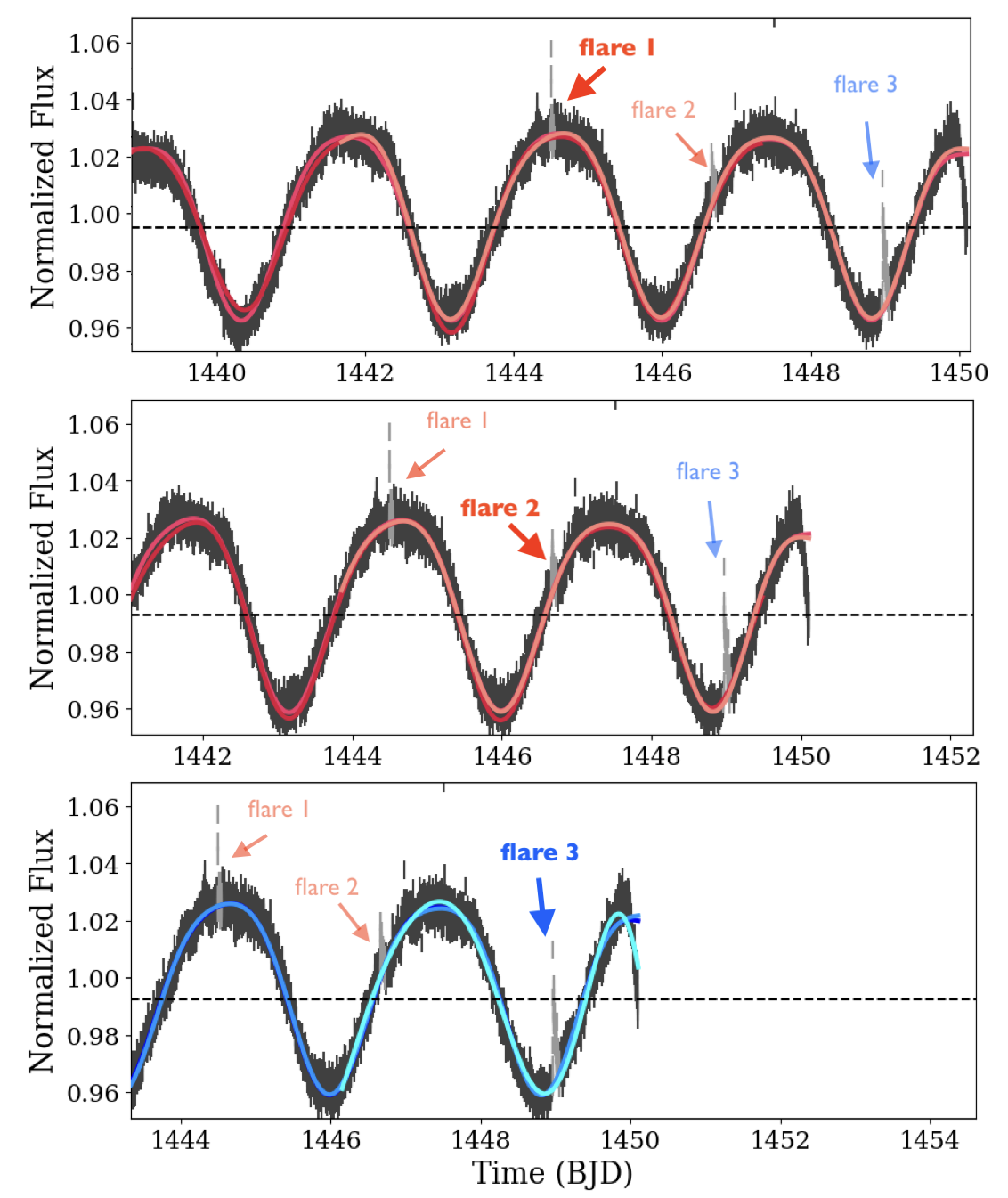}
    \caption{A set of three consecutive flares and their spot fittings. The fitted flare in question is centered and the window displays the segment of data used in its center fit. While fitting, the flare (in grey; see \ref{sec:fit_criteria}) is masked and only the surrounding measurements are considered. The modulation at flare time is then interpolated. Fits indicating positive spot modulation at the flare are shaded red; negative spot modulation, blue. Spot amplitude is calculated relative to the window median, which is denoted by a horizontal dashed line. The average normalized spot amplitude of the fits are 0.98 (top), 0.13 (middle), and -0.82 (bottom) respectively.} 
    \label{fig:pos_neg}
\end{figure}

\begin{figure}
    \centering
    \includegraphics[width=1\linewidth]{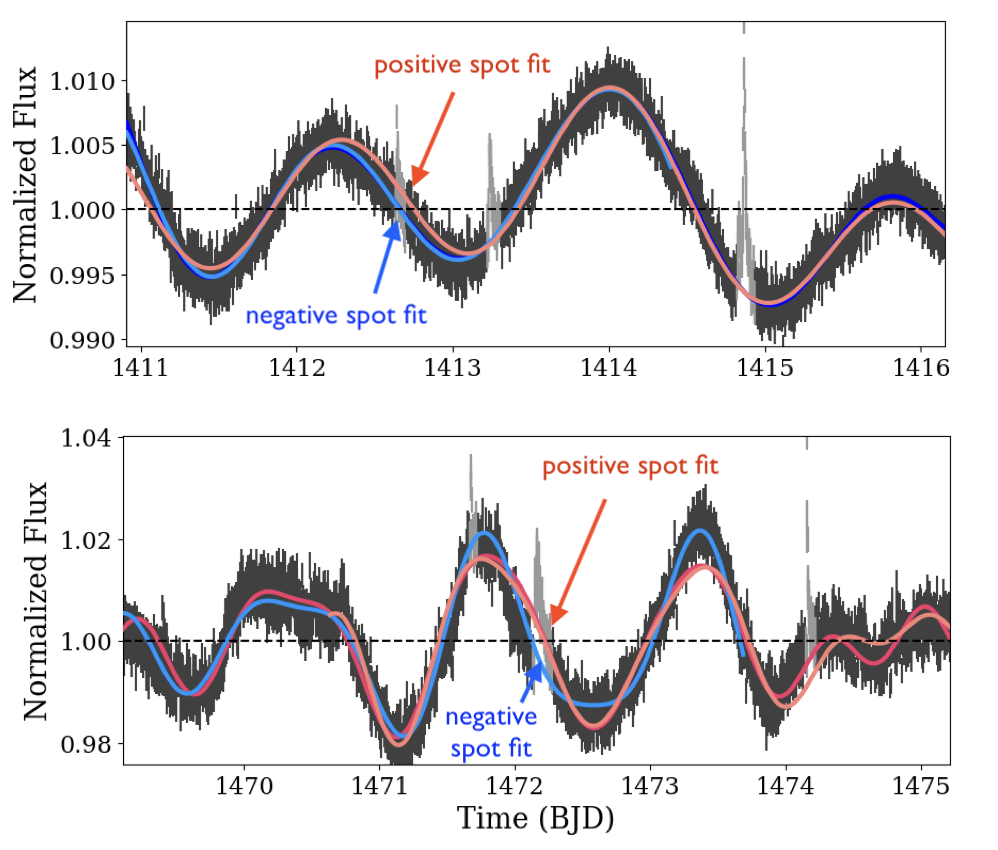}
    \caption{Two flares for which two fitted spot modulation curves disagree on whether the modulation at the flare is positive (in red) or negative (in blue), with light shades representing the rightmost fit window and dark shades for the leftmost fit window. This disagreement leads to the exclusion of the flare from our dataset. This vetting procedure is responsible for the V-shaped cutout around a normalized spot amplitude of 0 in the histogram in Fig.~\ref{fig:result}.} 
    \label{fig:disagree}
\end{figure}

To reduce the dependence of the fitted spot modulation on how the data window is defined, we also repeat the fit twice by shifting the data window left and rightwards by one local periodogram period respectively. If any fit shows disagreement in the sign of the spot modulation, the flare is discarded, as shown in Figure \ref{fig:disagree}. This cut means that we ignore flares that occur near the median spot flux because it is ambiguous if the star is slightly brigher or slightly dimmer than typical. We prefer to only include flares where it is clear that the star is brighter or dimmer.

The spot amplitude at flare peak of these three fits are then averaged, and the resulting value determines whether a flare is positive or negative (see Figure \ref{fig:pos_neg}).

\section{Filters and Bias Testing} \label{sec:bias}
In total we detected 19,422 flares from the \cite{2024Feinstein} sample, 14,541 flares from the \cite{2025Yudovich} sample, and 184,423 flares from the \cite{2025seli} sample.

Before comparing the relative occurrence of detected flares during periods of positive spot modulation as compared to negative spot modulation, there are biases that should be accounted for to ensure robust results. Various effects of the data collection from lightcurve detrending, flare detection, and spot modeling can bias the results towards a positive or negative correlation. Here we discuss the sources of such biases, how we addressed them in our samples, and a Monte-Carlo experiment we conducted to ensure the efficacy of the measures we employed.


\subsection{Poor Detrending} \label{sec:poor_detrend}

The detrending employed in this work is effective in treating long periodic behaviors seen from the changing number of spots in the line of sight of the viewer as a result of stellar rotation. However, more rapid or chaotic modulations in the lightcurve coming from rapidly spinning or pulsating stars prove more challenging. Lightcurves from these objects will not be sufficiently flattened to ensure quality flare detection, as sudden pulsations or significant residual spot modulation may be picked up as false-positive flares. Additionally, lightcurves coming from rapidly pulsating stars are not able to be captured by the spot modeling described in Section \ref{sec:double_D_sine}.

A lightcurve is expected to follow an underlying trend with scatter coming from the photometric error of the measured fluxes. If the spread of the flux points in the detrended lightcurve is significantly larger than the average photometric error, then the detrending method has not sufficiently flattened the lightcurve. The residual noise can significantly affect flare detection and signifies an inability to be followed by the spot modeling. 

We define a lightcurve to have been poorly detrended if $\frac{\sigma}{\text{flux err}} > 1.3$ where $\sigma$ is the standard deviation of the fluxes after detrending and \text{flux err} is the average photometric flux error. Across our samples, we saw 1,284 such lightcurves in the \cite{2024Feinstein} sample, 506 in the \cite{2025Yudovich} sample, and 10,500 in the \cite{2025seli} sample. These lightcurves were cut from analysis. This entailed cutting 4,253 flares from the \cite{2024Feinstein} sample, 4,202 from the \cite{2025Yudovich} sample, and 42,881 from the \cite{2025seli} sample.

\subsection{Poor Spot Modeling} \label{sec:poor_spot_mod}

The ability to determine whether or not a detected flare event occurs during periods of increased or decreased emission in a star depends on the ability of the fitted spot modulation to accurately track the trend in the lightcurve. If the best fit spot modulation poorly follows the trend in the lightcurve, the value of the fitted spot modulation at the time of the flare may be erroneous to the point that it does not accurately capture whether or not the underlying spot modulation flux is positive or negative. These poor fits can often be attributed to incomplete masking of the flare decay, which bias a fit towards an above average flux, or transits that bias in the opposite direction. 

For our study, we only consider spot modulation fits that had reduced $\chi^2$ values of less than 100 relative to the photometric flux to avoid catastrophic misfits that can affect whether or not a flare is accurately determined to have occurred during positive or negative spot modulation. Such a cut removed 404 flares from the \cite{2024Feinstein} sample, 163, from the \cite{2025Yudovich} sample, and 6,679 from the \cite{2025seli} sample.

\subsection{Low Amplitude Incompleteness} \label{sec:low_amp}
\begin{figure}
    \centering
    \includegraphics[width=0.48\textwidth]{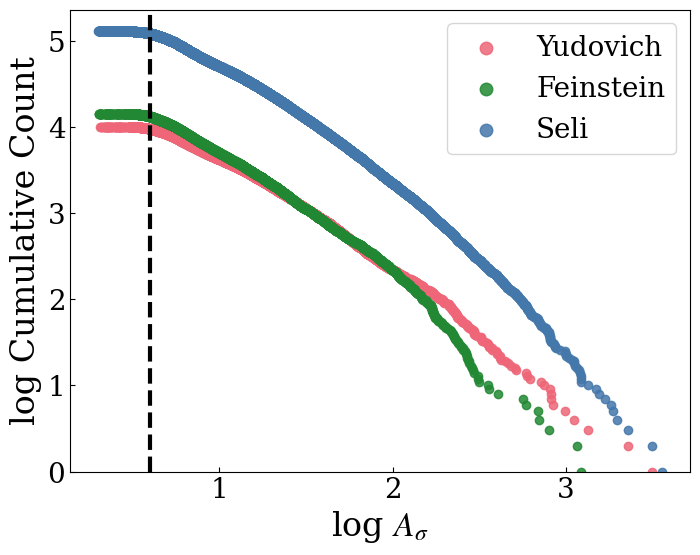}
    \caption{Cumulative number counts of flares in the \cite{2024Feinstein} sample demonstrating the number of flares of amplitude expressed in terms of the spread of the lightcurve, $\sigma$, $A_{\sigma}$, or higher expressed in log-space. Deviations from a straight line in log-log space are emblematic of incompleteness at low amplitude (left of the vertical dashed line) or small number statistics at high amplitude.}
    \label{fig:flare_completeness}
\end{figure}

Another potential source of bias in this study is incompleteness not only as a results of missing low amplitude flares with amplitudes $< 3\sigma$, but also the incompleteness due to residual spot modulation. Because of incomplete flattening of the light curve, there can be significant detection bias in favor of flares occurring at positive spot modulation, as they are more visible and easily distinguished compared to identical flares at negative spot amplitudes. Likewise, the incomplete masking of flares, particularly during flare decay as the light curve returns to being predominantly a function of spot modulation, can induce a similar positive bias. This incompleteness is particularly significant at low amplitudes and is visualized in Figure \ref{fig:flare_completeness}, showing the cumulative number counts of the flares based on their amplitudes in terms of $\sigma$. For the \cite{2024Feinstein} sample considered, the trend deviates from the expected straight line in log-space at $\log{A_{\sigma}} \approx 0.6$ which corresponds to an amplitude of $A_{\sigma} \approx 4$. 

This $4\sigma$ cut is consistent with the notion that, in lightcurves with residual red noise, $1\sigma$ will trace the amplitude of the modulation, thus adding it to the original threshold of $3\sigma$ will create a completeness criterion. Such a cut removed 1,072 flares from the \cite{2024Feinstein} sample, 626, from the \cite{2025Yudovich} sample, and 8,572 from the \cite{2025seli} sample.

\subsection{Spot Fit Disagreement} \label{sec:disagree}

Finally, we also do not consider any flare for which our spot modulation fits for the three windows around its peak disagree in whether the flare occurs at positive or negative modulation. This disagreement predominantly cuts flares which occur near median flux, as seen in Figure \ref{fig:disagree}. A slight difference in the fitted model parameters can change the sign of the spot modulation flux at the time of transit. We remove such systems since it is ambiguous if the star is more or less spotted than usual at the time of flare. The application of these cuts removed 1,749 flares from the \cite{2024Feinstein} sample, 856, from the \cite{2025Yudovich} sample, and 17,402 from the \cite{2025seli} sample. In the end we considered 11,944 flares from the \cite{2024Feinstein} sample, 8,694 flares from the \cite{2025Yudovich} sample, and 108,841 flares from the \cite{2025seli} sample to be from spotty stars and resistant to various biases.

\subsection{Flare Injection} \label{sec:synth_flares}
To test any implicit methodological and algorithmic bias in our flare detection and spot modulation fitting, we also conducted a flare injection and retrieval test on each sample. For each sector per star, flares retrieved by \textsc{toffee} are re-injected at random times using the quartic rise and double-exponential decay model from \cite{2014Davenport}, with FWHM equal to half of the found flare duration by \textsc{toffee} to account for incomplete detection.

Sections of the lightcurve containing the original detected flares are then masked according to \ref{sec:fit_criteria} – including any synthetic injected flares that coincide there as well. We then ran the flare detection of \textsc{toffee} on the lightcurves to test how many of the synthetic lightcurves we recover. For the \citet{2024Feinstein} sample, of the 19,422 flares injected, \textsc{toffee} recovered 17,183 flares, for a recovery rate of 88.5\%. After applying the cuts to poorly detrended lightcurves, poorly fit spot mods, and flares with amplitudes less than $4\sigma$, we returns a positivity of 51.05 $\pm$ 0.52\%. That is to say, in this injected sample, which should be perfectly 50/50, there is a slight bias towards finding flares at a positive spot modulation flux, at least for the \citet{2024Feinstein} sample. We applied the same injection-retrieval method to the \citet{2025Yudovich} and \citet{2025seli} samples, returning positivity values of $49.48\pm 0.63\%$ and $50.48\pm 0.17\%$ respectively.

Overall, \textsc{toffee} does not seem to have a strong systematic bias towards finding flares at positive or negative spot flux. Indeed, for the largest sample of flares - \citet{2025seli} - our bias test is compatable with 50\%, within $1\sigma$. Even though these biases seem minimal, we will use these injection-retrieval-derived positivity rates as the baseline in determining the true positivity rate for flares in Sect.~\ref{sec:results}. 


\section{Results \& Discussion}\label{sec:results}

\subsection{The Correlation Between Spots and Flares}
We applied \textsc{toffee} and subsequent spot-fitting to three existing flaring star catalogs compiled by \cite{2024Feinstein}, \cite{2025seli}, and \cite{2025Yudovich}. The original papers were used only to provide the stars - all flares and spots were independently discovered by \textsc{toffee}. The resulting distribution of spot amplitudes at flare times for each is shown below in Fig.~\ref{fig:result}.

We also conducted bootstrapping on each of the catalogs by resampling the normalized spot amplitudes of flares with replacement 100,000 times and recalculating the percentage of positive spot amplitudes. To calculate the significance of our result above the methodological bias tested in \ref{sec:bias}, we shifted the spot amplitude distributions being resampled to be centered at the spot amplitude percentile given by our injection-retrieval. 

\begin{figure}
    \centering
    \includegraphics[width=0.48\textwidth]{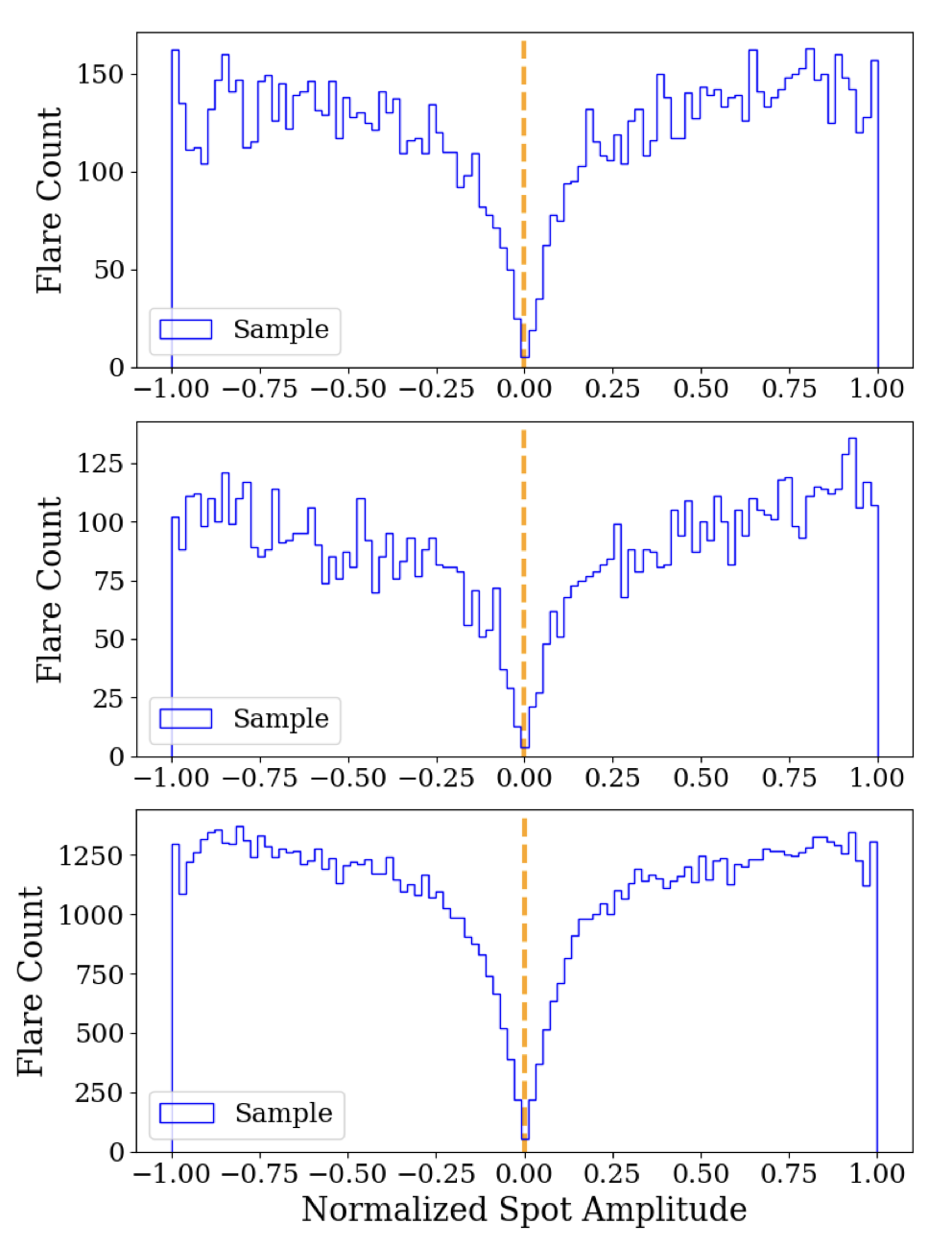}
    \caption{Normalized Spot Amplitude of TOFFEE Flares from Feinstein (top), Yudovich (middle), and Seli (bottom) Catalogs. The V-shaped cutout in the middle comes from our exclusion of flares for which the sign of the spot modulation fit was ambiguous (see Fig.~\ref{fig:disagree} and Sect.~\ref{sec:disagree})}
    \label{fig:result}
\end{figure}

Biases resulting from poor detrending, flare incompleteness, and poor spot modulation fitting (see Sect. \ref{sec:bias}) lead us to exclude 7,478 of 19,422 flares detected on the \cite{2024Feinstein} sample. Of the remaining 11,944, we find a positivity of 50.96 $\pm$ 0.46 \%. This is in agreement with the value of 51.05 $\pm$ 0.52 \% measured from the random injected sample.

On the \cite{2025Yudovich} sample, we filter 5,847 of 14,541 fitted flares, and obtain a positivity of 51.85 $\pm$ 0.54 \% of the remaining 8694, compared to 49.48 $\pm$ 0.63 \% measured from the random injected sample. Therefore, for the \citet{2025Yudovich} sample we find that flares occur more often when the stars is brighter (more spots), at 4.42 $\sigma$ significance.

On the \cite{2025seli} sample, we filter 75,534 of 184,423 fitted flares, and obtain a positivity of 49.71 $\pm$ 0.15 \% of the remaining 108,889, compared to  50.48 $\pm$ 0.17 \% measured from the random injected sample. This is a 5.07 $\sigma$ preference for flares to occur when the star is dimmer (more spots).

If we combine all three samples, we have 14,163 unique stars and the positivity rate is 49.97 $\pm$ 0.21\%. Otherwise put, across all tested stars and flares, there is no determined preferences for flares to occur when the star is brighter or dimmer.

\subsection{A Question of Faculae}

\begin{figure}
    \centering
    \includegraphics[width=0.48\textwidth]{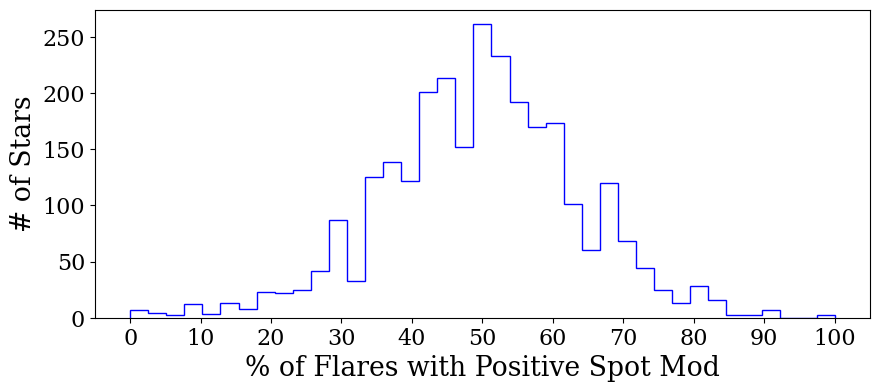}
    \caption{The percentage of flares that occur at positive spot modulation for the 2,750 spotty stars in the Seli Catalog containing at least 10 flares.}
    \label{fig:seli_by_tic}
\end{figure}

An objective statement from our study is that we find no statistically strong correlation between flare occurrence and the star being brighter or dimmer than average, throughout a star's rotation period. Throughout this paper we have considered the modulation of the stellar flux to be a direct representation of how many spots are facing the observer at a given time; a brighter flux means less spots and a dimmer flux means more spots. Indeed, this is how spots are treated in the vast majority of the literature.

However, the presence of dark spots is normally accompanied by brighter regions called faculae, complicating our interpretation. On the Sun spots are concentrated between $\pm 30^{\circ}$ of the equator. Faculae are located both near the spots and across a network that spans much more of the solar surface. The total surface area covered by faculae is $\sim 17$ times the area covered by spots \citep{Chapman1997}. However, faculae are a much lower contrast than spots. Consequently, despite a $17\times$ higher surface coverage, the increased flux from faculae only outweighs the flux deficit from spots by about $50\%$. Over the 11 year solar cycle, the Sun actually gets about 0.1\% brighter at solar maximum, because despite there being more spots, there are also more faculae, and this leads to a net brightening. \citet{Montet2017} studied long-term variations of Kepler-observed stars and found that slow rotators (like the Sun) are faculae-dominated, whereas fast rotators are spot-dominated, i.e. they get dimmer at the peak of the activity cycle when there are more spots present.

In our study we are studying flare occurrence relative to short term stellar variability, specifically stellar rotation. If increased spots mean that a star gets brighter due to increased faculae, our interpretation of a star's rotation curve could be inverted. That is to say, when the star gets darker during its rotation it could mean that \textit{less} spots have rotated into view, not more. 

Moreover, according to \citet{Shapiro2016}, the answer to this question depends on the spatial distribution of spots and faculae, the orientation of the observer and the observational bandpass used. For the Sun, at UV wavelengths faculae dominate the rotational brightness variability. At visible wavelengths, if the observer is aligned with the Solar equator (like we approximately are from Earth), then spots will typically dominate the rotational modulation. This is due to the higher contrast of spots and their location near the equator. When the observer is inclined relative to the Solar equator by more than $\sim 60^{\circ}$, then the rotational modulation at visible wavelengths is dominated by faculae. \citet{Canis2024} tested the combined effect of spots and faculae on other stars, showing that the presence of faculae can confuse the derived stellar rotation period, particularly if the star is observed close to pole-on.

For our study we are observing spots and flares in visible light with TESS. The relative orientation of the stellar spin axis could be a priori anything, although an exactly pole-on orientation would produce no rotational variability. We also cannot assume that the latitudinal distribution of starspots is the same on our sample (largely M-dwarfs) as it is on the Sun. Indeed, there is evidence that M-dwarfs predominantly have polar spots \citep{Strassmeier2009,Ilin2021}. Overall, faculae could influence or even dominate our rotation curves, confusing the implications of our results.

Our primary result is that the positivity rate of the spot modulation flux at the time of flare is $49.97\pm 0.21\%$. We interpret this as evidence that spots and flares are not correlated. Consider instead a hypothetical scenario where our sample is roughly an even mix between spot-dominated and faculae-dominated stars. In this scenario, if flares were indeed correlated with spot occurrence, then we would expect roughly half of the stars to have a low positivity rate, and half to have a high positivity rate, and hence the overall rate would still be $\sim 50\%$. We test this ambiguity by taking the largest sample of flares (Seli) and restricting it to stars with at least 10 flares, yielding a total of 2,750 stars. For each star we calculate the percentage of flares occurring at positive spot modulation. We then histogram this percentage across all applicable stars in Fig.~\ref{fig:seli_by_tic}. 

If all flares prefer heightened spots, but our sample contains a mixture of faculae-dominated and spot-dominated stars, then we would expect Fig.~\ref{fig:seli_by_tic} to be bimodal, with a peak at a low positivity rate and a peak at a high positivity rate. Instead, the results in  Fig.~\ref{fig:seli_by_tic} show that most stars have a roughly 50\% positivity rate. Overall, whilst there remains further investigation to be made into the effect of faculae, none of our analysis points to any correlations between flares and star spots.

\subsection{A Question of the Solar Comparison}

The majority of the stars in all three samples (\citealt{2024Feinstein,2025Yudovich,2025seli}) are M-dwarfs. The majority of flares discovered occur on M-dwarfs. Some of these M-dwarfs have flares more energetic than the famous Carrington event\footnote{An intense solar flare and associated geomagnetic storm of 1859. It remains the most energetic solar outburst ever directly recorded, producing widespread aurorae, major disruptions to telegraph systems and significant disturbances to the sleeping patterns of Colorado gold miners.}, yet on a daily basis. White light M-dwarf flares can cause the star to even double in brightness. It is worth noting that on the Sun the most energetic flares are the ones that are most strongly correlated with the presence of spots \citep{2014Guo,Li2023,2023Oloketuyi}. However, overall the majority of the flares used in our statistical analysis are different to Solar flares, both in terms of absolute energy and relative brightness increase.

\subsection{Other Sources of Confusion}

It is likely that some of our ``spotted single stars'' are actually tight non-eclipsing binaries, with flux modulations not from spots but instead due to ellipsoidal, rotation or Doppler beaming effects \citep{Faigler2012}. Such effects are only relevant for binaries with periods less than 10 days, which corresponds to the tail of the binary separation distribution \citep{Raghavan2010}, and hence this false positive is probably a small contribution to our sample. Some of our sample could also be pulsating single stars (e.g. $\gamma$ Doradus and $\delta$ Scuti variables). However, our sample contains very few stars above 6500 K, and such stars rarely produce observable white light flares. Finally, the large 21 arcsecond TESS pixels mean that contamination between multiple stars on the same pixel is possible, and hence there may be cases where the spots and flares are actually coming from different stars but we attribute them to a common origin.

\section{Conclusion}\label{sec:conclusion}

We applied a new flare detection algorithm \textsc{toffee} to discover 228,663 flares on 16,305 stars across 3 samples of flaring stars compiled in the literature. The majority of the stars are M-dwarfs. We discovered a subset of 14,163 stars with clear photometric modulation attributable to star spots. From theses stars we calculated the spot modulation flux of the star at the time of 218,386 flares. We implemented numerous checks and filters to ensure that our final flare sample was not biased as a function of the spot modulation flux. We discovered that flares occurred when the star was brighter than normal  $49.97\pm 0.21\%$. That is to say, we found no correlation between the occurrence of flares and starspots. Whilst this lack of correlation has been seen in earlier studies, our study includes improvements in bias correction, a simpler yet more rigorous test, and a much larger sample of stars and flares. 

On the Sun, solar flares typically occur near sunspots, and hence there is a strong correlation between sunspots and solar flares. This is seemingly not a universal stellar phenomenon.


\section*{ACKNOWLEDGMENTS}

This work began as a class project in AST-51/151 ``Astrophysics Laboratory'' taught in Spring 2024 and Spring 2025 at Tufts University. We appreciate the hard work of all of the students throughout these two semesters. Martin congratulates the Spring 2025 students for killing the \citet{2024Martin} proposed relationship between spots and flares on CM Draconis, through the addition of new TESS data.

\bibliography{Bibliography}{}
\bibliographystyle{aasjournal}



\end{document}